\documentclass[%
reprint,
superscriptaddress,
 amsmath,amssymb,
 aps,
]{revtex4-1}

\usepackage{graphicx}
\usepackage{dcolumn}
\usepackage{bm}
\usepackage[version=4]{mhchem}
\usepackage{siunitx}
\usepackage{todonotes}
\usepackage{xcolor}

\begin{document}


\title{Spectroscopy of $^{33}$Mg with knockout reactions}


\author{D. Bazin}
\email{bazin@nscl.msu.edu}
\affiliation{National Superconducting Cyclotron Laboratory, 640 S. Shaw Lane, East Lansing, MI 48824-1321}
\affiliation{Department of Physics and Astronomy, Michigan State University, East Lansing, MI 48824-1321, United States}
\author{N. Aoi}
\affiliation{RIKEN Nishina Center, RIKEN, 2-1 Hirosawa, Wako, Saitama 351-0198, Japan}
\affiliation{Research Center for Nuclear Physics, 10-1 Mihogaoka, Ibaraki, Osaka 567-0047, Japan}
\author{H. Baba}
\affiliation{RIKEN Nishina Center, RIKEN, 2-1 Hirosawa, Wako, Saitama 351-0198, Japan}
\author{J. Chen}
\affiliation{National Superconducting Cyclotron Laboratory, 640 S. Shaw Lane, East Lansing, MI 48824-1321}
\author{H. Crawford}
\affiliation{Nuclear Science Division, Lawrence Berkeley National Laboratory, Berkeley, California 94720, USA}
\author{P. Doornenbal}
\affiliation{RIKEN Nishina Center, RIKEN, 2-1 Hirosawa, Wako, Saitama 351-0198, Japan}
\author{P. Fallon}
\affiliation{Nuclear Science Division, Lawrence Berkeley National Laboratory, Berkeley, California 94720, USA}
\author{K. Li}
\affiliation{RIKEN Nishina Center, RIKEN, 2-1 Hirosawa, Wako, Saitama 351-0198, Japan}
\author{J. Lee}
\affiliation{RIKEN Nishina Center, RIKEN, 2-1 Hirosawa, Wako, Saitama 351-0198, Japan}
\affiliation{Department of Physics, University of Hong Kong, Pokfulam Road, Hong Kong, China}
\author{M. Matsushita}
\affiliation{RIKEN Nishina Center, RIKEN, 2-1 Hirosawa, Wako, Saitama 351-0198, Japan}
\author{T. Motobayashi}
\affiliation{RIKEN Nishina Center, RIKEN, 2-1 Hirosawa, Wako, Saitama 351-0198, Japan}
\author{H. Sakurai}
\affiliation{RIKEN Nishina Center, RIKEN, 2-1 Hirosawa, Wako, Saitama 351-0198, Japan}
\affiliation{Department of Physics, University of Tokyo, 7-3-1 Hongo, Bunkyo, Tokyo 113-0033, Japan}
\author{H. Scheit}
\affiliation{RIKEN Nishina Center, RIKEN, 2-1 Hirosawa, Wako, Saitama 351-0198, Japan}
\affiliation{Technische Universit\"at Darmstadt, 64289 Darmstadt, Germany}
\author{D. Steppenbeck}
\affiliation{RIKEN Nishina Center, RIKEN, 2-1 Hirosawa, Wako, Saitama 351-0198, Japan}
\author{R. Stroberg}
\affiliation{Department of Physics, University of Washington, Seattle, WA 98195-1560, USA}
\author{S. Takeuchi}
\affiliation{RIKEN Nishina Center, RIKEN, 2-1 Hirosawa, Wako, Saitama 351-0198, Japan}
\author{H. Wang}
\affiliation{RIKEN Nishina Center, RIKEN, 2-1 Hirosawa, Wako, Saitama 351-0198, Japan}
\author{K. Yoneda}
\affiliation{RIKEN Nishina Center, RIKEN, 2-1 Hirosawa, Wako, Saitama 351-0198, Japan}
\author{C.X. Yuan}
\affiliation{Sino-French Institute of Nuclear Engineering and Technology, Sun Yat-Sen University, Zhuhai 519082, China}





\date{\today}

\begin{abstract}
The structure of \ce{^33Mg} was investigated by means of two knockout reactions, one-neutron removal from \ce{^34Mg} and one-proton removal from \ce{^34Al}. Using comparative analysis of the population of observed excited states in the residual $^{33}$Mg, the nature of these states can be deciphered. In addition, the long-standing controversy about the parity of the \ce{^33Mg} ground state is resolved using momentum distribution analysis, showing a clear signature for negative parity. Partial cross section measurements are compared with the results of eikonal reaction theory combined with large-scale shell model calculations of this complex nucleus located in the island of inversion, where configuration mixing plays a major role.
\begin{description}
\item[PACS numbers]
\end{description}
\end{abstract}

\maketitle


\section{Introduction}

The structure of the nucleus \ce{^33Mg} has been the subject of much debate over the past few years. It is located in the so-called island of inversion \cite{WAR90} where intruder configurations arise due to the quenching of the N=20 shell gap. Numerous experimental evidence point to this region as strongly deformed, from the lowering of the first 2$^+$ excited states in \ce{^32^,^34^,^36Mg} \cite{DET79,YON01,GAD07,DOO13,MIC14} and more recently \ce{^32Ne} \cite{DOO09,MUR19} as well as large B(E2) transition strengths in \ce{^31Na} \cite{PRI00} and \ce{^30Ne} \cite{MIC14,DOO16}.

The origin of the deformation in this region is now well established and revealed from the evolution of the effective single-particle energies leading to the disappearance of the N=20 shell gap in neutron-rich isotopes, and the resulting enhancement of multi-particle multi-hole excitations across the narrowed gap. 
Shell model calculations can now reproduce the narrowing of the gap between the $1d_{3/2}$ and ($1f_{7/2}$, $2p_{3/2}$) orbitals, and the appearance of intruder states \cite{OTS20}. In a mean-field picture \cite{HAM12}, this deformation can be understood in terms of a degeneracy of the $1f_{7/2}$ and $2p_{3/2}$ orbits.

The case of \ce{^33Mg} however remains less clear. $\beta$-decay measurements from \ce{^33Na} seemed to indicate a spin-parity of $3/2^+$ based on log(ft) values \cite{NUM01}, that was interpreted as a 1p-1h excitation across an inversion of the $1d_{3/2}$ and $1f_{7/2}$ orbitals. This interpretation was further reinforced by a Coulomb excitation experiment where the observed state at 485 keV was assigned a spin-parity of $5/2^+$ \cite{PRI02}. On the other hand, the measured negative magnetic moment of \ce{^33Mg} is in direct contradiction with this interpretation, suggesting a spin-parity of $3/2^-$ and a 2p-2h excitation \cite{YOR07}. A year after the magnetic moment measurement, results on the $\beta$-decay from \ce{^33Mg} heated the debate again by reporting a large branching ratio to the $5/2^+$ ground state of \ce{^33Al}, and tried to reconcile the overall picture by proposing a mixing of 1p-1h and 3p-3h configurations \cite{TRI08,YOR10,TRI10}. A subsequent measurement of the inclusive momentum distribution in the one-neutron knockout from \ce{^33Mg} indicated a large occupation of the $2p_{3/2}$ orbital, an indication of its lowering \cite{KAN10}. Although a fit of spectroscopic factors assuming a $3/2^-$ ground state seemed to be closer to Monte-Carlo Shell Model predictions, no conclusion on the spin-parity was reached in that paper. A Coulomb dissociation experiment \cite{DAT16} unveiled some evidence of multiparticle-hole ground state configuration involving the $1s_{1/2}$ and $1p_{1/2}$ orbitals. Finally, excited states of \ce{^33Mg} were populated in the fragmentation reaction of a \ce{^46Ar} radioactive beam, and detected using the GRETINA $\gamma$-ray tracking array \cite{PAS13}. The energies of the populated states were measured, and from the energy differences the presence of a rotational band based on a $3/2^-$ ground state was inferred \cite{RIC17}.

In this work, the ground and excited states configurations of \ce{^33Mg} are investigated by means of one-neutron and one-proton knockout reactions from \ce{^34Mg} and \ce{^34Al}, respectively. The selectivity of these two reactions plays an important role in the identification of the final states populated in \ce{^33Mg}. From the $0^+$ ground state of \ce{^34Mg}, only one partial wave can contribute to the direct feeding a given excited state in \ce{^33Mg}, therefore an analysis of the momentum distribution shape reveals the orbital angular momentum, and by deduction, the parity of that state. In contrast, the removal of a proton from \ce{^34Al} can proceed via a number of partial waves, depending on the spin-parity of its ground state, but as this valence proton most probably occupies the $1d_{5/2}$ orbital, the population of final states in \ce{^33Mg} is expected to be more selective and favor states of the same parity as the ground state of \ce{^34Al} for a $\ell$=2 proton removal.

\section{Experiment}

The experiment was conducted at the Radioactive Ion Beam Factory (RIBF) located on the RIKEN campus of Tokyo, Japan, and operated jointly by RIKEN and the Center of Nuclear Study from the University of Tokyo. The \ce{^34Mg} and \ce{^34Al} radioactive beams were produced from the fragmentation of a 345 MeV/u \ce{^48Ca} primary beam on a 15 mm thick \ce{^9Be} target. The radioactive beams were selected and filtered by the BigRIPS fragment separator \cite{KUB12} up to the F8 focal plane, where the 1032 mg/cm$^2$ \ce{^9Be} reaction target was placed. Surrounding this target was the NaI(Tl) $\gamma$-ray array DALI2 \cite{TAK14}, which recorded the Doppler-shifted $\gamma$-rays emitted by the reaction residues. The $^{33}$Mg residues produced in the knockout reactions were then selected and collected by the ZeroDegree Spectrometer (ZDS) operated in dispersive mode in order to measure their momentum. The incoming energies of the \ce{^34Mg} and \ce{^34Al} radioactive beams were 242.5 MeV/u and 229.6 MeV/u, and their average intensities 4.5$\times$10$^3$ pps and 6.5$\times$10$^4$ pps respectively.

One of the main goals of this experiment was to measure partial and inclusive cross sections populating the various states of the \ce{^33Mg} residue, therefore careful calibrations of the incoming fluxes of the \ce{^34Mg} and \ce{^34Al} projectiles were performed by setting the ZDS on the magnetic rigidity of these projectiles after slowing down in the target, and measuring the ratios of \ce{^34Mg} and \ce{^34Al} detected at the focal plane of the ZDS relative to the total number of particles counted by the plastic scintillator located upstream of the target at the F7 focal plane of BigRIPS. The purities deduced from these measurements are 69\% and 60\% for \ce{^34Mg} and \ce{^34Al}, respectively. In addition, the acceptance and momentum dispersion of the ZDS were measured by scanning the unreacted \ce{^34Mg} beam across the focal plane of the ZDS by varying its magnetic rigidity. The momentum dispersion was determined to be 4.5 cm/\%, close to the expected value from optics calculations. 

\begin{figure}[h]
\centering
\includegraphics[width=\columnwidth]{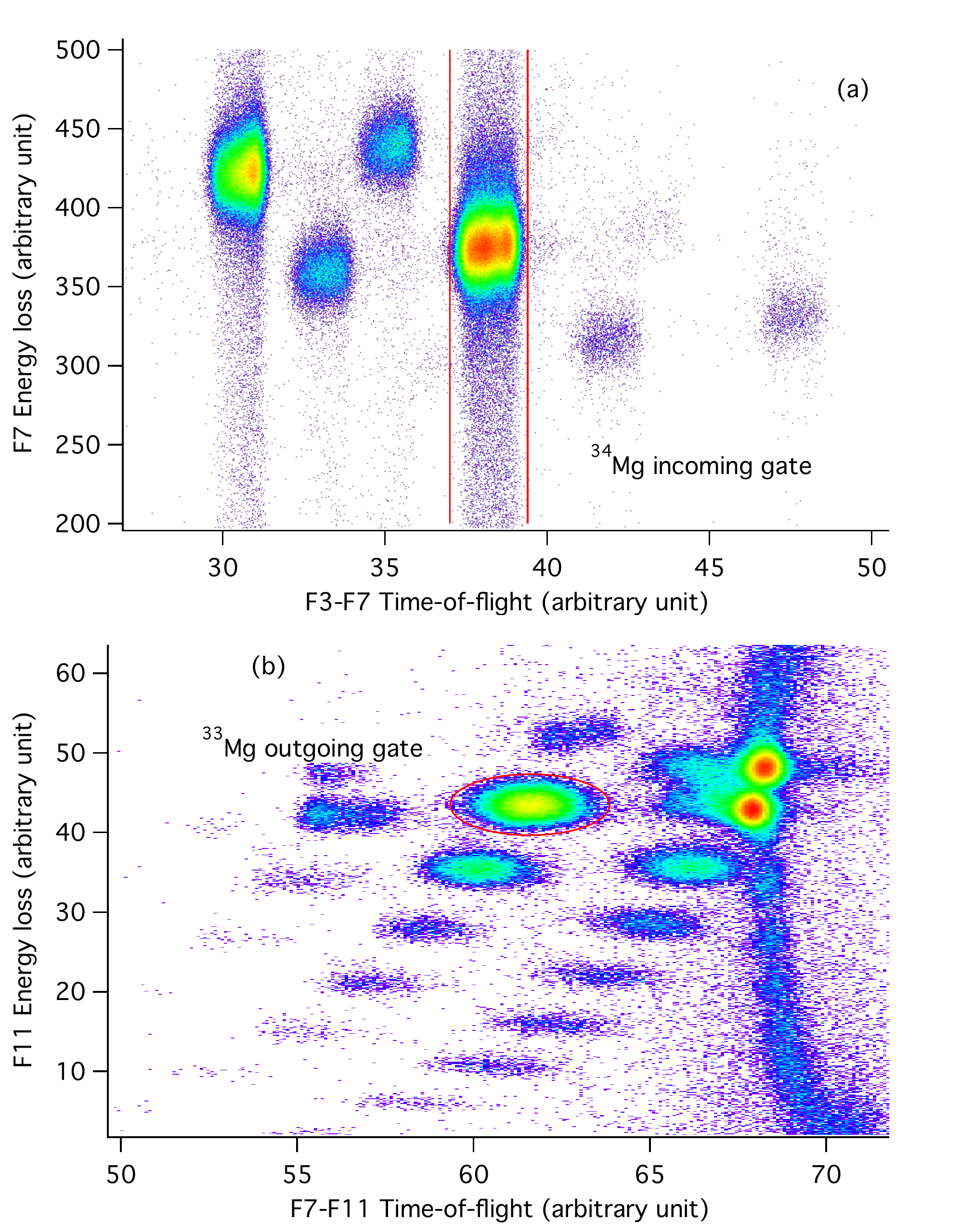}
\caption{Particle identification spectra using time-of-flight versus energy loss before (a) and after (b) the reaction target. The plots are shown for the case of the \ce{^34Mg} radioactive beam. Plot (b) is gated on the \ce{^34Mg} incoming gate shown in plot (a). The double structure observed at the \ce{^34Mg} location corresponds to inelastic scattering of incoming \ce{^34Mg} beam events through the Be target. Due to the dispersive mode of the ZDS, these particles are transmitted on the high side of the momentum acceptance, and hit the F11 scintillator close to its edge near where the light guide is located and produce different energy losses.}
\label{pid}
\end{figure}

The particle identification prior to the reaction target was performed using the detectors located at F3 and F7 of the BigRIPS fragment separator. The time-of-flight measured between the plastic scintillators at F3 and F7 was used in combination with the energy loss signal from the F7 ion chamber. However, because the contaminants were clearly separated in time-of-flight from the \ce{^34Mg}/\ce{^34Al}, and there was significant pile-up in the ion chamber at higher rate, the incoming \ce{^34Mg} gate was limited to the time-of-flight only. The momentum width of the incoming radioactive beam was set to \SI{0.5}{\percent} using the momentum slits located at F1.

The reaction residue particle identification after the secondary target was done using the detectors of the F11 focal plane of the ZDS. Since the ZDS was set in dispersion mode, the time-of-flight between the F7 and F11 plastic scintillators was corrected for the trajectory length dependency using the position in the dispersive direction measured at F11. The energy loss was measured by the F11 ion chamber. Fig. \ref{pid} shows the particle identification spectra before (a) and after (b) the reaction target in the case of the \ce{^34Mg} beam.

The parallel momentum of the residues was deduced from the dispersive position measured at the F11 focal plane and the measured dispersion of \SI{4.5}{cm/\percent}. The broadening effect due to the momentum width of the incoming beam, was canceled using the dispersive position measurement at the F5 focal plane. A resulting momentum resolution of \SI{0.1}{\percent} was obtained, dominated by the energy straggling in the reaction target and the intrinsic resolution of the position detectors.
Owing to the large forward momentum of the projectiles and their resulting \ce{^33Mg} residues, the 4\% momentum acceptance of the ZDS was large enough to collect all residues with negligible losses. 

\section{Results}
The quantities measured during this experiment were the parallel momentum distribution of the \ce{^33Mg} residues, the energy of the $\gamma$-rays emitted following nucleon removal, and the cross sections of the reactions. Due to the large background coming from the breakup of the Be reaction target and the Bremsstrahlung radiation, the $\gamma$-ray multiplicity M recorded by the DALI2 array was very large, and some of the $\gamma$-ray transitions could only be observed in the M=1 spectra. For this reason, the M=1 and M=2 spectra were used to identify the transitions and possible $\gamma$-$\gamma$ coincidences. However, because the Geant4 simulations are not able to simulate the background, the intensities of the $\gamma$-ray transitions were fitted using the M$\leq$6 spectra, the multiplicity of 6 being determined as the value beyond which no significant increase in intensities was observed.

\subsection{One-neutron knockout from \ce{^34Mg}}
The inclusive cross section of this reaction was measured using the incoming beam normalization method described in the previous section. The cross section obtained is 93(2) mb, where the error bar is dominated by the systematic errors associated with the determination of the number of incoming \ce{^34Mg} particles. 

\begin{figure}[h]
\centering
\includegraphics[width=\columnwidth]{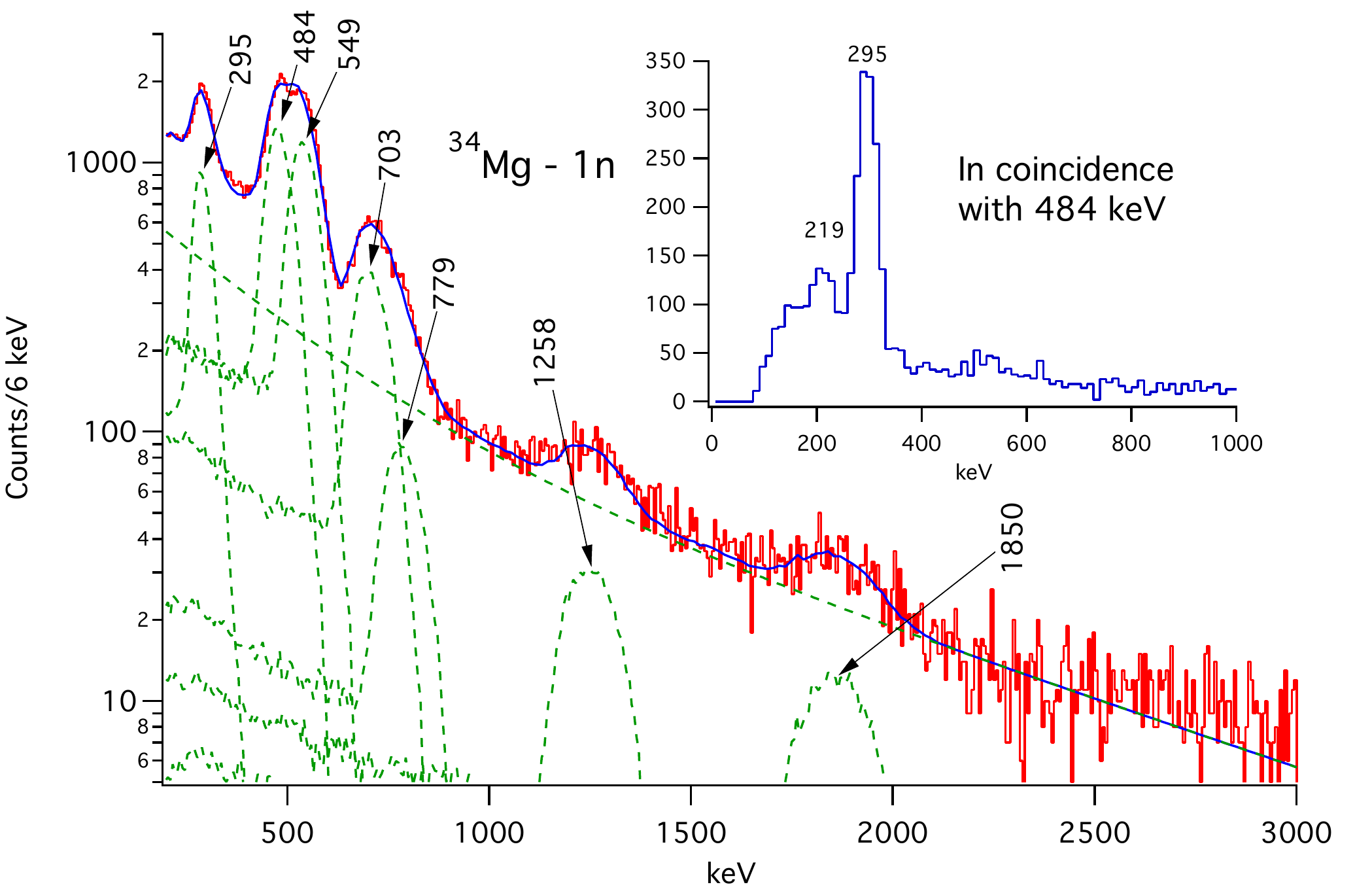}
\caption{M=1 $\gamma$-ray spectrum observed in coincidence with \ce{^33Mg} residues produced from one-neutron knockout off \ce{^34Mg} projectiles. Seven $\gamma$-ray transitions are identified, and their energies determined from fitting the spectrum with calculated line-shapes from Geant4 simulations and a double-exponential background (all shown in dashed lines). The inset shows the $\gamma$-$\gamma$ coincidence spectrum gating on the 484 keV transition where a large coincidence is observed with the 295 keV $\gamma$-ray line, as well as a weaker transition at 219 keV.}
\label{g34Mg1n}
\end{figure}

The Doppler-corrected $\gamma$-ray spectrum measured in coincidence with the one-neutron knockout of \ce{^34Mg} is shown in Fig. \ref{g34Mg1n}. This spectrum was generated using an add-back procedure to reduce the peak-to-background ratio, and gated on multiplicity one (M=1) to further enhance the identification of peaks. 

Eight transitions were observed and their energies fitted using simulated line-shapes from a Geant4 simulation of the array. 
The inset of Fig. \ref{g34Mg1n} shows the $\gamma$ spectrum in coincidence with the 484(6) keV transition. A very clear coincidence is observed between the 484(6) keV and 295(7) keV transitions, in agreement with the reported observations in \cite{NUM01} and \cite{RIC17}. The weaker 219(8) keV transition corresponding to the 221 keV transition observed in \cite{NUM01} and \cite{RIC17} could only be resolved in the coincidence spectrum due to the large Bremsstrahlung background present at this beam energy. Although its presence cannot be ruled out because of the limited energy resolution, the 759 keV transition reported in \cite{NUM01} in coincidence with the 484 keV transition is only hinted in this neutron knockout data. A transition at 1857 keV was reported in \cite{NUM01} but not assigned to any of the $\beta$-decay daughters of \ce{^33Na}. The energy however is very close to the 1850(40) keV transition observed in our data, so it likely corresponds to the same transition (the one-neutron separation energy of \ce{^33Mg} is 2.07 MeV).

\begin{figure}[h]
\centering
\includegraphics[width=\columnwidth]{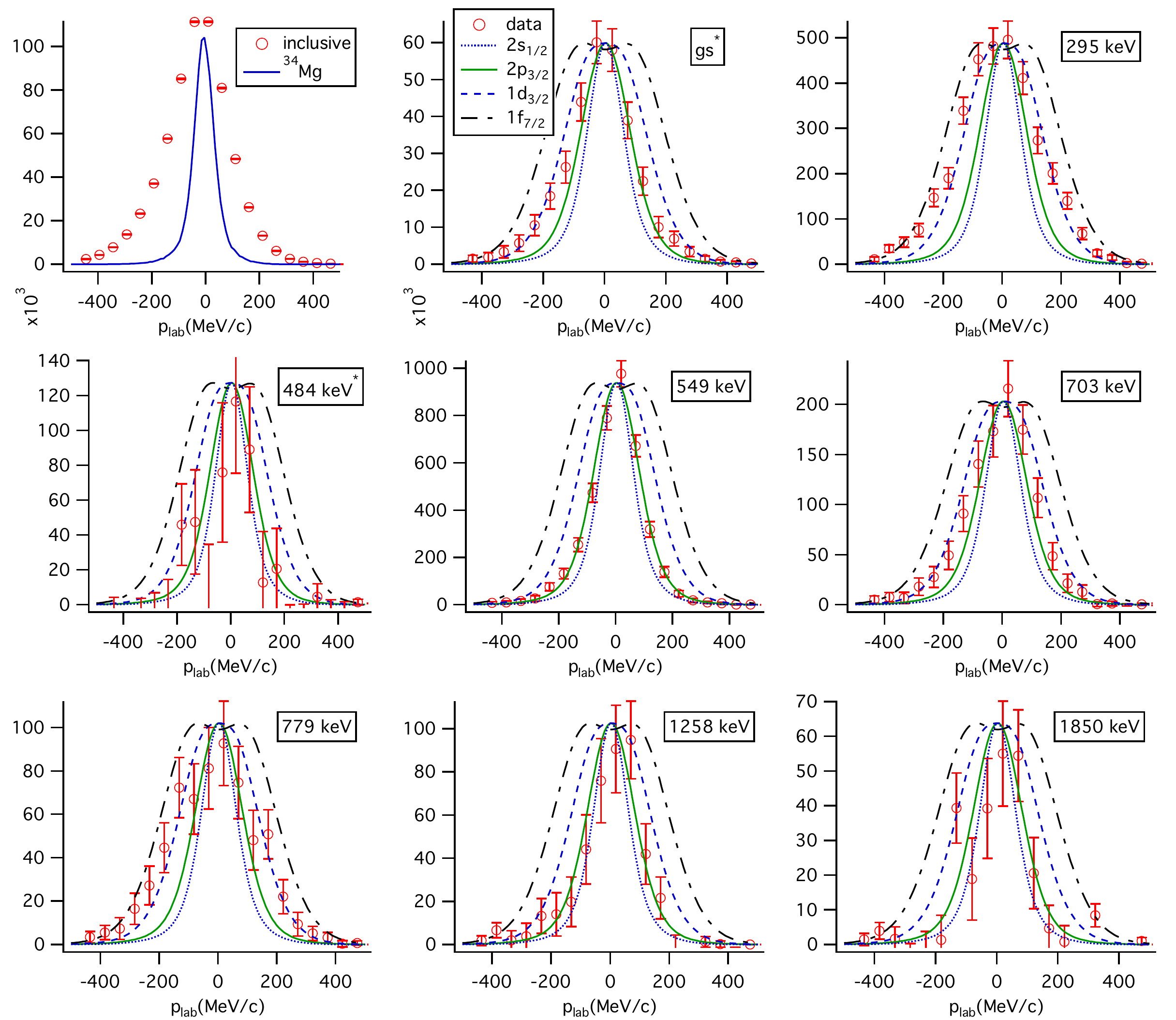}
\caption{\ce{^33Mg} residues momentum distributions observed in coincidence with $\gamma$-ray transitions, each labeled with the corresponding energy. The top left panel show the inclusive momentum distribution as well as the unreacted \ce{^34Mg} distribution obtained with the ZDS spectrometer centered on its energy after the reaction target. The latter illustrates the momentum resolution obtained in this experiment with the ZDS spectrometer set in dispersive mode. Unlike the momentum distributions gated on the $\gamma$-ray peaks from the M=1 spectrum, the ground state and 484 keV distributions (indicated by an asterisk) are obtained by subtracting the contributions from their feeding states, using the $\gamma$-ray spectrum for all multiplicities and the branchings deduced for each of the feeder states. The distributions are compared to calculations from the eikonal model (see text).}
\label{pparall}
\end{figure}

The M=1 $\gamma$-ray spectrum was used to extract the momentum distributions corresponding to the populated excited final states in \ce{^33Mg}. Due to the strong overlap between some of the peaks, the shapes of the momentum distributions may be cross-contaminated. Also, the background subtraction could only be performed using the data around 1 MeV and above, since at lower energies the spectrum is dominated by the numerous transitions and their Compton contributions. In order the minimize the cross-contamination where peaks overlap, the gates used to extract the momentum distributions were shifted from the centroid of the peak away from the contaminating peak. The resulting momentum distributions are displayed in Fig. \ref{pparall}, each labeled with the transition energy.

The top left panel of the figure shows the inclusive momentum distribution, while the top middle panel shows the momentum distribution extracted for the direct feeding of the ground state. This distribution is calculated by subtracting the contributions from the excited feeder states weighted from the branchings deduced from the $\gamma$-ray spectrum without multiplicity cut. This procedure is necessary because of the inability to accurately simulate the $\gamma$ background in the Geant4 simulations. The observed cascades between excited states are also taken into account in the subtraction. The same procedure is used to deduce the momentum distribution corresponding to the direct feeding of the 484 keV state, although the error bars are large because of the strong feeding from the 779 keV state (see section \ref{spins} below).

Each experimental momentum distribution is compared to calculated distributions using the eikonal model \cite{HAN03}, assuming the removed neutron belongs to the orbitals $2s_{1/2}$, $2p_{3/2}$, $1d_{3/2}$, or $1f_{7/2}$. To convert them into the laboratory reference frame, these distributions are scaled by the relativistic factor $\gamma$=1.243 corresponding to the velocity of the \ce{^34Mg} projectiles at mid-target. For most distributions, the angular momentum of the removed neutron is clearly identified by comparing the calculated shapes to the experimental data. The ground state momentum distribution, although displaying the usual low momentum tail usually observed in knockout reactions, is best reproduced by the $2p_{3/2}$ distribution. This establishes without ambiguity the negative parity of the ground state of \ce{^33Mg}, in agreement with the magnetic moment measurement \cite{YOR07} and conclusions from the inclusive measurement of neutron knockout from \ce{^33Mg} \cite{KAN10}.

The shapes of the momentum distributions associated with other transitions are identified, with the exception of the one associated with the 484 keV state for which the error bars are large due to the feeding subtraction, and the one associated with the 703 keV for which both p-wave and d-wave components seem to be present. This latter observation is most likely due to cross-contamination from the 779 keV d-wave component due to the proximity of the $\gamma$-ray transitions, and indicates that the momentum distribution associated with the 703 keV transition is most likely a p-wave. Due to the limited statistics, the shapes associated with the higher energy transitions at 1258 keV and 1850 keV are also ambiguous between a s-wave and p-wave assignment, especially since the difference between the two theoretical distributions is relatively small and their widths not much larger than the resolution.
Noteworthy is the non-observation of any transition associated with an $\ell$=3 momentum distribution. This is investigated in the next section.

\subsection{Search for the $\ell$=3 strength}
\label{l3}
The parallel momentum distributions shown in fig. \ref{pparall} clearly exclude the observation of any $\ell$=3 strength associated with any of the prompt $\gamma$-rays detected in coincidence with the $^{33}$Mg residues, whereas the lowest 7/2$^-$ state corresponding to the 0p0h configuration is expected within all shell model calculations to be populated with a significant strength. This surprising result could be explained if this state is long-lived and therefore decays after the residues have left the sensitive area of the $\gamma$-ray array. 

\begin{figure}[h]
\centering
\includegraphics[width=\columnwidth]{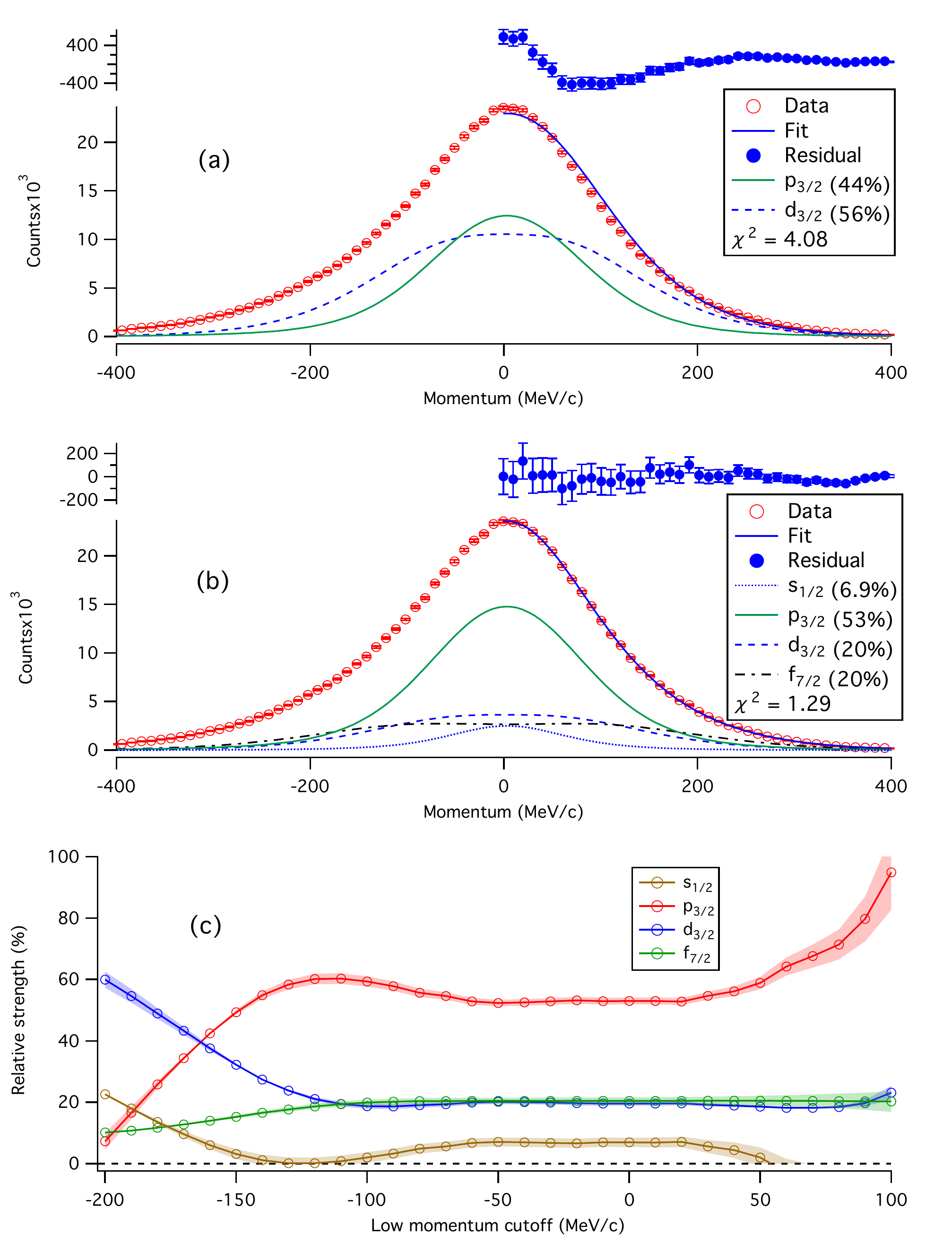}
\caption{Fits to the inclusive momentum distribution of one-neutron knockout from \ce{^34Mg}. The top panel (a) shows the results of a fit using only $\ell$=1 and $\ell$=2 calculated momentum distributions, whereas the middle panel (b) shows the same fit adding $\ell$=3 and $\ell$=0 components. Panel (c) shows the evolution of the various $\ell$ components with respect to the low momentum cutoff in the fit. See text for details.}
\label{fitinclusive}
\end{figure}

In order to test this hypothesis and evaluate the branching ratio of this missing strength, two fits of the inclusive parallel momentum distribution were performed, using calculated momentum distributions for $\ell$=1,2 and $\ell$=0,1,2,3, respectively. The inclusive momentum distribution has a much larger statistics because it is not subject to the $\gamma$-ray array detection efficiency. The results of the fits are shown in fig. \ref{fitinclusive}.
All momentum distributions are centered on the mean momentum in the laboratory frame. The only fitting parameters are the strengths of each component, and only the high momentum portion of the data is used. This restriction is necessary because the calculated momentum distributions from the eikonal model are symmetrical, whereas the measured ones show an asymmetry due to the low momentum tail. This effect is well known and originates from the lack of energy conservation in the eikonal model \cite{HAN03}, as well as dissipative contributions from inelastic excitations of the core residue \cite{STR14}. The reduced $\chi^2$ values (displayed on the figure) indicate a much better agreement with the experiment in the second fit (b). The residuals obtained in the first fit (a) indeed clearly indicate that a third component with a shape similar to the $\ell$=3 distribution is needed.  The $\ell$=0 component is added for completeness, as it should be present as well in the inclusive data, although it is not distinguishable from the $\ell$=1 component due to the resolution of the parallel momentum measurement. In addition, the percentages that represent the relative strengths of each component show that the first fit (a) gives almost equal strength for the $\ell$=1 and $\ell$=2 components, which is in contradiction with the results obtained from the $\gamma$-ray branching ratios in the previous section, where the $\ell$=1 component clearly dominates. 

The relative strength of the f$_{7/2}$ component is 20$\pm$2\%, where the error bar is determined by varying the upper bound of the fitting region, as well as the validity test explained in the next paragraph. The isomeric nature of the 7/2$^-$ state makes it impossible to subtract its component from the inclusive distribution in order to extract the ground state momentum distribution, as is done in fig. \ref{pparall}, therefore it is composed of both the ground and isomer states components. Indeed, a fit to this ground state momentum distribution using both p-wave and f-wave components also gives a much better match to the data, although its accuracy is limited by the relatively large error bars. Based on the 49$\pm$1\% branching ratio for the ground and isomer states deduced from the observed $\gamma$-ray intensities, this fit gives a relative strength of 15$\pm$5\% for the f-wave component, which is compatible with the 20$\pm$2\% determination based on the inclusive distribution fit. However, it should be noted that this 49\% is likely an upper limit because some of the weaker $\gamma$-ray transitions are probably missed due to the limited resolution. A reduction of the deduced intensity to the ground and isomer states would reconcile better the two f-wave relative strength determinations outlined above. Because of its much smaller error bars, the determination based on the inclusive momentum distribution is kept as the final result.

To further test the validity of this extraction of the $\ell$=3 strength, the evolution of the various $\ell$ components returned by the fit shown in (b) is plotted as a function of the low momentum cutoff (c). A clear plateau is observed between cutoff values of -60 MeV/c and 20 MeV/c, indicating a robust determination of the various components. The shaded area around the curves shows the evolution of the errors on the relative strength determination.

\subsection{One-proton knockout from \ce{^34Al}}
\begin{figure}[h]
\centering
\includegraphics[width=\columnwidth]{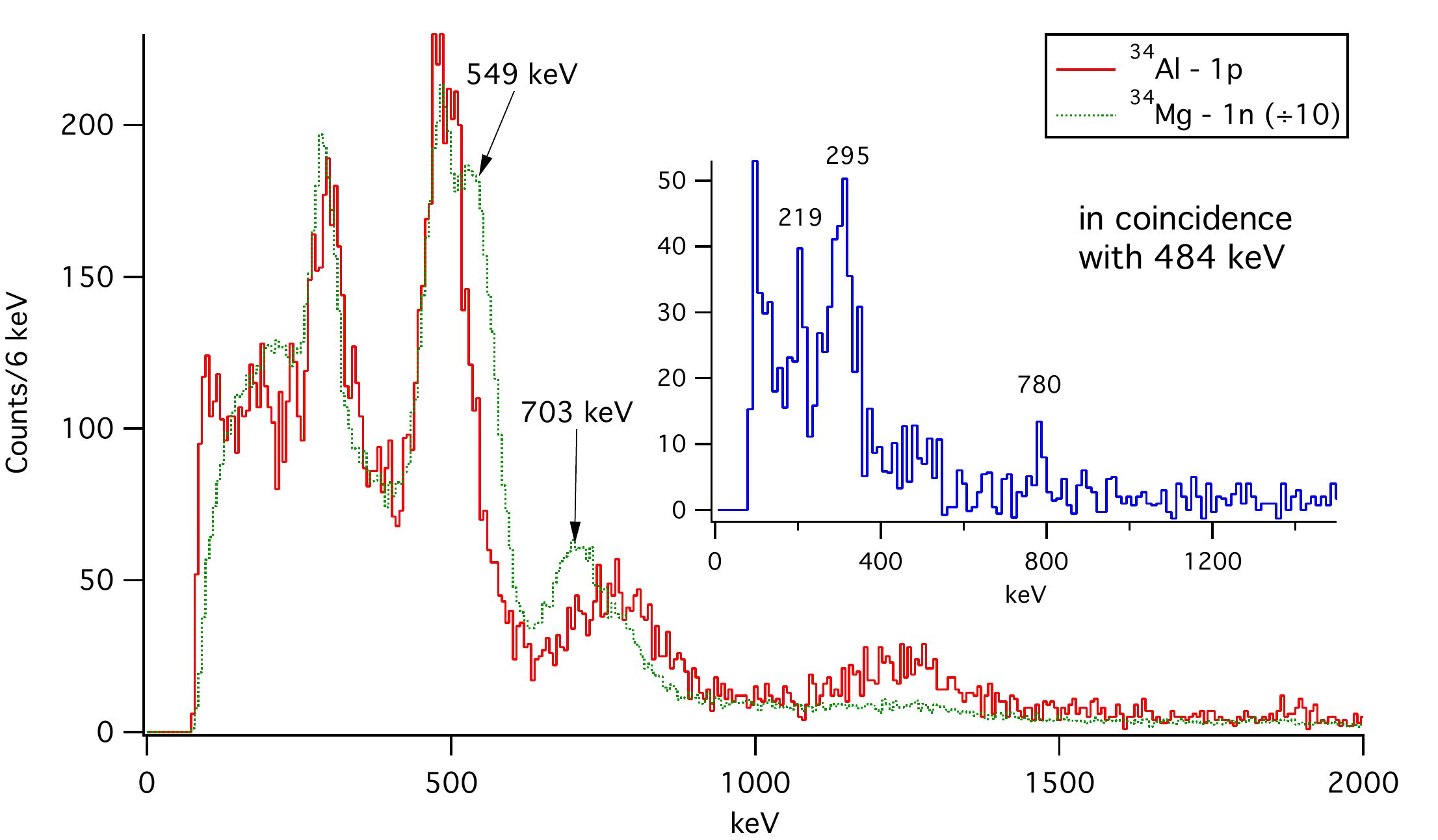}
\caption{M=1 $\gamma$-ray spectra from one-proton knockout off \ce{^34Al} (solid red) and one-neutron knockout off \ce{^34Mg} (dashed green) reduced by a factor 10. The difference in population of transitions between the two reactions is clearly illustrated. The inset shows the $\gamma$-$\gamma$ coincidence spectrum gated on the 484 keV transition. In addition to the already observed coincidence with the 219 keV and 295 keV transitions, an additional weaker coincidence is observed at around 780 keV, although it is not visible in the coincidence spectrum from the neutron knockout data.}
\label{pknock}
\end{figure}

The $\gamma$ spectrum observed in coincidence with \ce{^33Mg} residues produced from the one-proton knockout of \ce{^34Al} are shown in Fig. \ref{pknock}, superimposed on the spectrum observed from the one-neutron knockout of \ce{^34Mg}. Although the limited energy resolution of the DALI2 array doesn't allow to resolve all the transitions, the comparison between the two spectra clearly shows the disappearance of the 549 keV and 703 keV transitions in the one-proton knockout data. Unlike in the one-neutron knockout reaction, the population of final states in \ce{^33Mg} from this reaction involves several partial waves because the ground state of the odd-odd \ce{^34Al} is not a 0$^+$, and is confirmed to be a 4$^-$ with a close-by 1$^+$ isomeric state at 47 keV with a half-life of 21.6(15) ms \cite{LIC17,HAN17}. It likely that both the normal 4$^-$ and intruder 1$^+$ long-lived states are populated in the fragmentation reaction used to produce \ce{^34Al} from \ce{^48Ca}, although it is not known in what proportions. For these reasons, the momentum distributions of the \ce{^33Mg} residues from this reaction are not as valuable as in the neutron knockout case and were not analyzed.

The proton knockout reaction will preferably populate proton-hole states in \ce{^33Mg}, which are likely to correspond to the removal of a proton from the $1d_{5/2}$ orbital. However, because both fundamental and isomeric states of \ce{^34Al} are likely present in the incoming beam, it is not possible to use the selectivity of this reaction to pinpoint the parity of the states populated.

Nevertheless, there are clear differences between the two observed $\gamma$-ray spectra. The two transitions that are clearly suppressed in the proton knockout data compared to the neutron knockout are the 549 keV and the 703 keV, that both correspond to the feeding of negative parity states, as identified from the momentum distributions of the \ce{^34Mg}-1n reaction. This could indicate, as suggested by \cite{NUM01}, that the 549 keV transition originates from the state at 703 keV and feeds the unobserved 7/2$^-$ isomeric state, which would then be placed at 154 keV. This hypothesis is however at odds with the non-observation of the 549 keV transition in \cite{RIC17} whereas the 703 keV is clearly populated by that reaction.

The coincidence spectrum shown in the inset of fig. \ref{pknock} reveals the same cascades going through the 484 keV state as in the neutron knockout data, with the addition of a weak coincidence peak at around 780 keV. The energy sum of this third cascade is 1264 keV, very close to the 1258 keV transition observed in both neutron and proton knockout data. The number of counts in the region of the 1258 keV transition is much larger in the proton knockout data than in the neutron knockout data, which could explain why this cascade is not visible in the latter coincidence spectrum. It is however surprising that the negative parity state observed at 1258 keV in the neutron knockout data would be more strongly populated in the proton knockout reaction. Unfortunately the limited resolution of the $\gamma$-ray array does not allow to determine whether another close-by state is present in this region. For this reason, the possible cascade transition at 780 keV is ignored in the level scheme presented below.

The inclusive cross section measured for this reaction is 3.1(2) mb.

\section{Discussion}
\subsection{Spin assignments and branching ratios}
\label{spins}
\begin{figure}[h]
\centering
\includegraphics[width=\columnwidth]{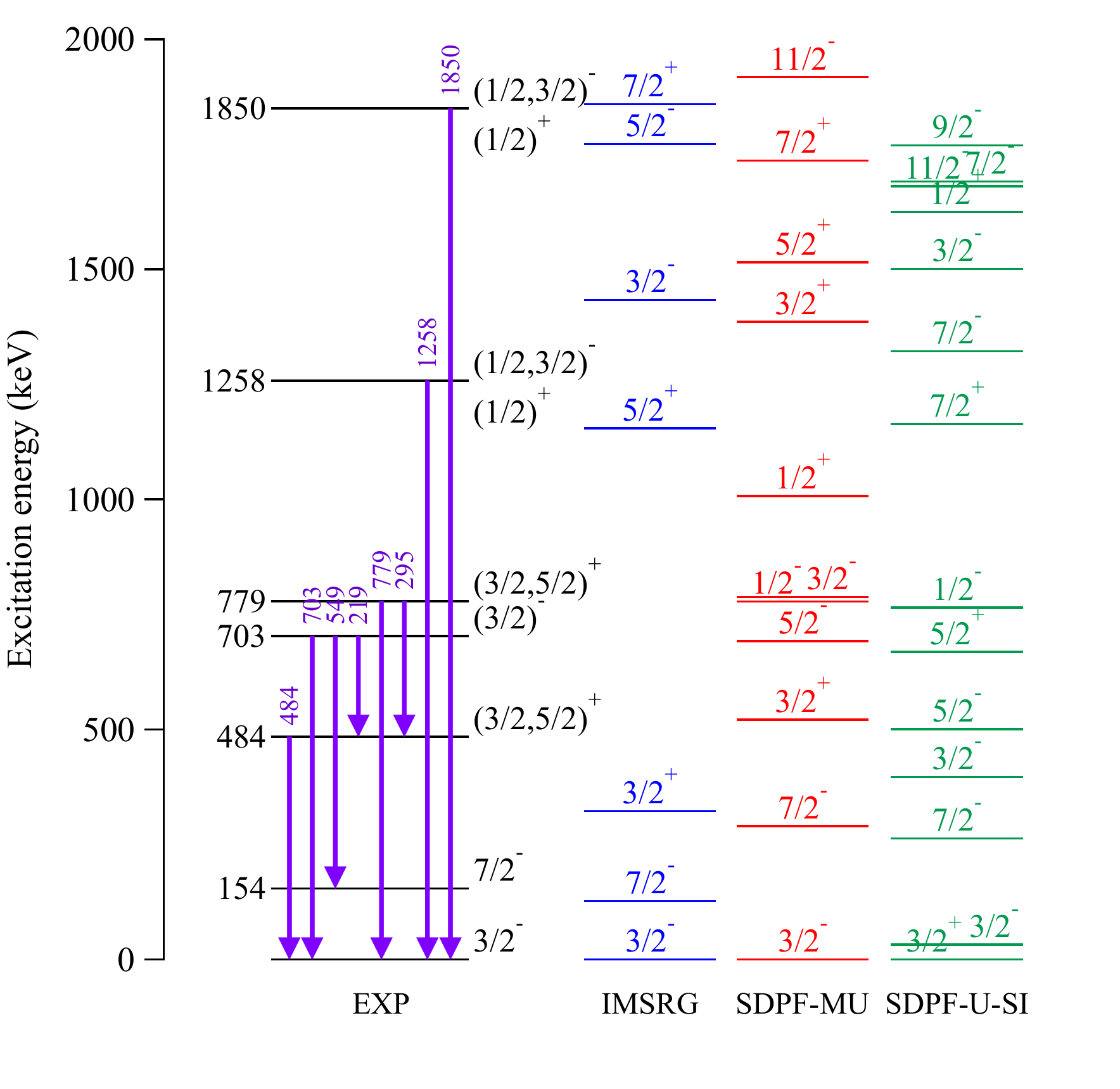}
\caption{Experimental and calculated level schemes for \ce{^33Mg}. The experimental parity assignments are based on the shapes of the momentum distributions observed in coincidence with the transitions. The tentative spin assignments are chosen from comparisons with earlier works as well as shell model calculations. The calculated level schemes use the various shell model interactions described in the text.}
\label{scheme}
\end{figure}

Based on the results and analysis presented above, a level scheme corresponding to the best hypothesis extracted from this data is presented in fig. \ref{scheme}. The transitions observed in this work are indicated by the arrows. The placement of the 7/2$^-$ state at 154 keV is based on indirect evidence only, with the assumption that this state has a long lifetime that prevented its decay from being detected as a prompt $\gamma$-ray.

The data clearly indicates that the state at 779 keV has positive parity, and likely correspond to either the $3/2^+$ or $5/2^+$ state calculated from shell model. This result is in contradiction with the hypothesis that this state is part of a rotational band built on the $3/2^-$ ground state, as reported in \cite{RIC17}. The strong feeding to the 484 keV state, also observed in \cite{RIC17}, indicates that the 484 keV has likely also a positive parity, as also suggested by shell model calculations (see section \ref{shell} below). However, the present data does not allow a firm assignment due to the large error bars and fluctuations in the momentum distribution obtained after subtracting the feeding. 

The three states at 703 keV, 1258 keV and 1850 keV that have negative parity from their associated momentum distributions, could correspond to the negative parity states calculated from the shell model, although it is not possible from the low level of statistics to firmly assign an angular momentum of either $\ell$=0 or 1 from the momentum distributions in coincidence with the 1258 keV and 1850 keV transitions.

The results showing intensities, branching ratios and partial cross sections are summarized in Tab. \ref{table1}. The table is ordered by increasing level energy, with feeding transitions from each level listed immediately below the ground state transition. One complication arises from the non-observation of the isomeric state in the $\gamma$-ray branching ratio analysis, for which the branching ratio is included in the ground state. For this reason the column labeled $b_\gamma$ differs from the final branching ratio $b$ for transitions involving the isomeric and ground states.

\begin{table}
\centering
\begin{tabular}{ccccccc}
\hline
$E_{level}$(keV) &	$E_\gamma$(keV)	&	$I_\gamma$(\%)	&	$b_\gamma$(\%)	&	$b$(\%) & $\sigma$(mb) & $\ell$ \\
\hline
0		&				&			&	49(1)	&	29(3)   &   27(4)	&   1\\
154 	&	unobserved  &	    	&	    	&	20(2)	&   18(4)   &   3\\
484		&	484(6)		&	23(1)	& $<$4(1)	& $<$4(1)   & $<$3(1)   &  (2)\\
703		&	703(8)		&	7(2)	& $>$22(3)	& $>$22(3)	& $>$20(3)  &   1\\
    	&	549(7)		&	15(1)	&	    	&	        &           &   1\\
		&	219(8)		&			&			&			&           &    \\
779		&	779(12)		&	4(1)	&   23(1)	&    23(1)  &    21(1)  &   2\\
		&	295(7)		&	19(1)	&			&			&           &   2\\
1258	&	1258(15)	&	1.4(5)	&   1.4(5)	&   1.4(5) &    1.3(5) & 0,1\\
1850	&	1850(40)	&	0.8(5)	&	0.8(5)	&	0.8(5)  &   0.7(5)	& 0,1\\
\hline
\end{tabular}
\caption{Levels, $\gamma$-ray energies, intensities, deduced branching ratios and partial cross sections. The $b_\gamma$ branching ratios result from the $\gamma$-ray intensity analysis, whereas the $b$ branching ratios take into account the presence of the isomeric level at 154 keV, which is speculative and not directly observed. The last column indicates the $\ell$-value observed from the gated momentum distributions, as well as the fit to the inclusive momentum distribution.}
\label{table1}
\end{table}

Under the assumption that the first 7/2$^-$ state is mostly populated and isomeric, the relative strength deduced in sec. \ref{l3} corresponds to unobserved strength in the branching ratios deduced from the $\gamma$-ray analysis presented in Tab. \ref{table1}, where it is assigned to the ground state. Based on that assumption, the branching ratios for the ground state and the 7/2$^-$ state are revised to 29$\pm$3\% and 20$\pm$2\%, corresponding to partial cross sections of 27(4) mb and 18(4) mb, respectively.

Due to the large background at low energy and its relative low intensity, the 219 keV transition observed in coincidence with the 484 keV transition is not visible in the singles spectrum and could not be included in the global fit of $\gamma$-ray intensities, therefore its intensity is unknown. Since this transition originates from the 703 keV level, the resulting branching ratios and partial cross sections for the 484 keV and 703 keV levels are taken as lower and upper limits, respectively. Therefore, only the largest feeding contribution from the 295 keV transition has been subtracted to obtain the momentum distribution corresponding to the 484 keV level displayed in fig. \ref{pparall}.

\subsection{Comparison to shell model}
\label{shell}
The partial cross sections are compared to calculated ones using spectroscopic factors from different shell model interactions. The single-particle cross sections are calculated using the eikonal model \cite{HAN03} relevant at these energies. The usual prescriptions for the sizes of the bodies used in these calculations are followed: the \ce{^9Be} target nucleus is modeled with a Gaussian density distribution of width 2.36 fm, while the projectile and residue nuclei densities are calculated from Hartree-Fock calculations using the SkX force. 

The shell model interactions used to calculate the spectroscopic factors are a IMSRG-derived interaction with an \ce{^20O} core, and 2 interactions using an \ce{^16O} core with 0-5$\hbar\omega$ excitations from sd to pf shell for valence neutrons. Valence protons are constrained in the sd shell only, because their cross-shell excitations are less important in the neutron-rich Mg isotopes. Also, the inclusion of such excitations demands much larger calculation resource.

The IMSRG interaction was generated by the VS-IMSRG method with ensemble normal ordering, as described in \cite{STR17}, using the Magnus formulation \cite{MOR15}. The input Hamiltonian is the 1.8/2.0(EM) interaction described in \cite{HEB11}, evaluated in an oscillator basis frequency hw=16 MeV, with truncations $e \equiv 2n+\ell \leq e_{max}=12$ and $e_1+e_2+e_3 \leq E_{3max}=14$. A Gloeckner-Lawson \cite{GLO74} center-of-mass term $\beta_{cm}(H_{cm} -3/2\hbar\omega)$ is added to push spurious states out of the spectrum.
For more details on the treatment of the center of mass in this context, see \cite{MIY20}. The resulting valence space interactions were diagonalized using the code NuShellX \cite{BRO14}. To compute the spectroscopic factors, we do not consistently-evolve the $a^{\dagger}$ operator, (work on implementing this is in progress). Since the spectroscopic factors are only used for a qualitative comparison, we expect this to be sufficient in the present context. When computing the spectroscopic factor, we perform all calculations using the interaction derived with the $^{34}$Mg reference.

The SDPF-MU \cite{UTS12} and SDPF-U-SI \cite{NOW09} interactions are widely used Hamiltonians for the sdpf region. They are both constructed to describe the properties of neutron-rich Si isotopes. It is therefore reasonable to use them to study $^{33}$Mg and nearby nuclei.

\begin{figure}[h]
\centering
\includegraphics[width=\columnwidth]{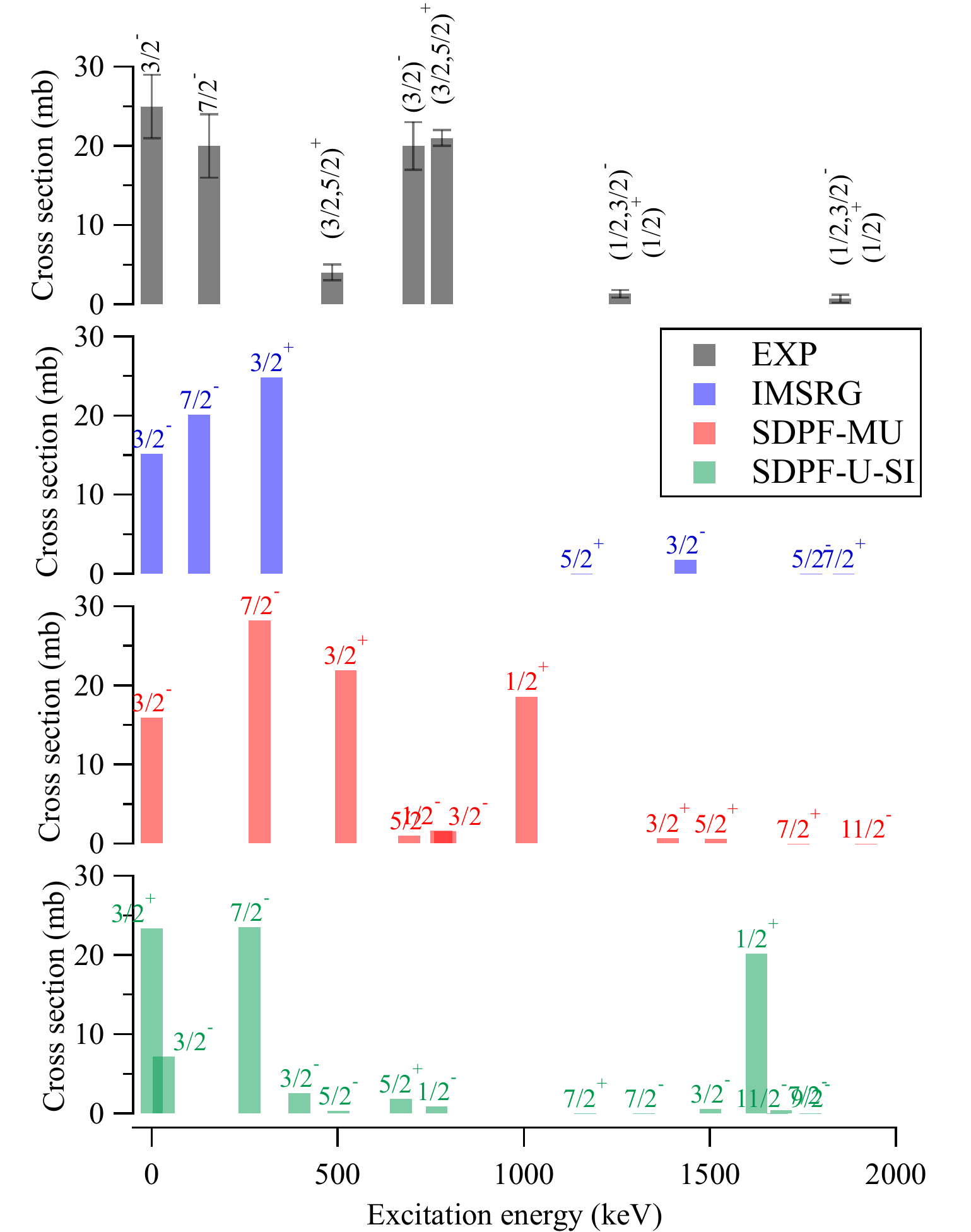}
\caption{Comparison between the measured and deduced partial cross sections and calculations, for the one-neutron knockout reaction from \ce{^34Mg}. The calculated cross sections are deduced from the shell model spectroscopic factors using various interactions, and the eikonal model single-particle cross sections.}
\label{partial}
\end{figure}

A spin-parity assignment of $3/2^+$ of the 484 keV state would imply that the Coulex experiment \cite{PRI02} performed on $^{33}$Mg measured an E1 transition. In this paper however, the assumptions were that the $J^{\pi}$ of the gs and 484 keV state are reversed. The B(E1) they deduced from the Coulex cross section is 0.035(10) $e^2$fm$^2$ using that assumption. Even though the parity assignment used in \cite{PRI02} is incorrect, the deduced B(E1) is unchanged by swapping the parities of the two states, because the spins of both inital and final states are the same, and the correction factor for a transition going in the opposite direction is (2J+1)/(2J'+1). This large value of the B(E1) is similar to the one observed in \ce{^27Ne} \cite{LOE18}, and could indicate similar effects of core excitation and deformation as the source of this large dipole strength. The shell model calculation based on the SDPF-MU interaction gives a much smaller value of 5.06e-6 e$^2$ fm$^2$, but it is normally difficult for shell model to reproduce B(E1) values in a limited model space.

The SDPF-MU and IMSRG calculations seem to best match the data, with the correct level ordering and qualitative reproduction of partial cross sections. It is clear however that the low-lying levels of \ce{^33Mg} and nearby nuclei in the island of inversion are very sensitive to the local single particle energies, namely, the effective single particle energies (ESPE). These were recently explored in nearby isotopes using a newly developed interaction  \cite{TSU17}, that showed the extent of ph mixing in the configuration of \ce{^31Mg} low-lying states, and the role that three-nucleon and tensor forces play in the evolution of the ESPE. It would be interesting to compare  calculations using this new interaction to the results presented here.

The non-observation of a 154 keV $\gamma$-ray in this experiment corresponding to the decay from the inferred 7/2$^-$ state to the ground state is likely due to two factors: the high background at low energy due to Bremsstrahlung radiation, and the possible isomeric nature of this state. Some estimates of the lifetime from shell model calculations follow. The 7/2$^-$ is not a very pure 0h0p state, especially in SDPF-MU. In SDPF-U-SI, the probability of pure $\pi$(d5/2)$^4$ $\nu$(f7/2)$^1$ configuration is 32.44\%, while in SDPF-MU, it is just 12.82\%, where several 2p2h $\nu$(f7/2)$^1$ configurations contribute. The B(E2) values are 88.4 and 74.5 e$^2$ fm$^4$ (effective charges e$_p$=1.5, e$_n$=0.5) for SDPF-MU and SDPF-U-SI, respectively, and the corresponding half-lives are 100 and 119 ns. With the IMSRG interaction, the prediction gives a lifetime of 180 ns assuming an energy of 150 keV and B(E2) of 10 Weisskopf unit, which is close to the value obtained for \ce{^31Mg} \cite{SEI11}. These estimates indicate that it would be impossible to detect this transition as a prompt $\gamma$-ray, and challenging as a delayed one because the lifetime is similar to the time-of-flight from the reaction target to a possible decay station at the end of the ZDS.

Finally the inclusive cross sections measured for both one-neutron and one-proton knockout reactions allow to determine the so-called reduction factor \cite{TOS14} that relate them to theoretical cross sections calculated from shell model spectroscopic factors and single-particle cross sections from the eikonal model. The interaction used in the shell model calculations is SDPF-MU, which seems to match the present experimental data the best. The \ce{^34Mg} to \ce{^33Mg} asymmetry energy is $\Delta$S = S$_n$-S$_p$ = -18.19 MeV. A theoretical cross section of 135.6 mb is obtained by summing all partial cross sections to individual states calculated in the shell model. With the inclusive cross section of 93(2) mb measured in this experiment, the reduction factor obtained is 0.69(2). For the one-proton knockout the situation is complicated by the fact that the relative population of the 1$^+$ isomer in the \ce{^34Al} beam is unknown. Nevertheless, the theoretical cross sections calculated using either the 4$^-$ or 1$^+$ are very close, 11.61 mb and 12.54 mb, respectively. The \ce{^34Al} to \ce{^33Mg} asymmetry energy is $\Delta$S = S$_p$-S$_n$ = 12.65 MeV. From the measured inclusive cross section of 3.1(2) mb, a reduction factor of 0.26(2) in obtained where the average value of the theoretical cross section is used, and the error bar takes into account the experimental error and difference between the theoretical values. These two reduction factor determination fall within the systematics presented in \cite{TOS14}, although somewhat on the low side of the observed band. They show once again a strong reduction of the cross section when a deeply bound nucleon is removed, compared to a calculation using an independent-particle model.

\section{Conclusion}
In this work we have studied the structure of \ce{^33Mg} by means of one-neutron and one-proton removal reactions from radioactive beams of \ce{^34Mg} and \ce{^34Al}, respectively. Momentum distributions in coincidence with prompt $\gamma$-rays recorded at the reaction target were analyzed and revealed the parity assignment of the populated final states in \ce{^33Mg}. The ground state momentum distribution obtained by subtraction from the inclusive momentum distribution is compatible with a p-wave shape, thereby confirming the 3/2$^-$ spin assignment of the \ce{^33Mg} ground state without ambiguity.

The state at 779 keV previously assigned as part of a rotational band built on top of the 3/2$^-$ ground state \cite{RIC17} has a positive parity, an observation in contradiction with this hypothesis. The strong feeding to the 484 keV state points to a similar parity, and comparison with shell model calculations give tentative spin assignments of 3/2$^+$ and 5/2$^+$. The states observed at 703 keV, 1258 keV and 1850 keV on the other hand have negative parity. The spin assignments for those states becomes more difficult in part due to the limited resolution of the Doppler-reconstructed $\gamma$-ray measurements, and the inability to distinguish between $\ell$=0 and $\ell$=1 shapes due to the finite resolution of the momentum reconstruction.

With a spin-parity assignment of the 484 keV state to 3/2$^+$, the strength measured in the Coulomb excitation experiment \cite{PRI00} would correspond to an E1 transition, with a B(E1) of 0.035(10) $e^2fm^2$. This large electric dipole strength is similar to the one observed in \ce{^27Ne} \cite{LOE18}, although the weak binding and low angular momentum conditions are not as extreme as in the case of \ce{^33Mg}. This may indicate that strong deformation and core excitations play an important role in this nucleus.

The absence of observation of any $\ell$=3 momentum distribution, which would reveal the feeding of the normal 0p0h configuration 7/2$^-$ state, is surprising. This likely indicates that this state has an isomeric nature, and therefore decayed too far from the $\gamma$-ray detection array. This assumption is further reinforced by the fitting of the inclusive momentum distribution which indicate that an $\ell$=3 component is needed. A fit including the $\ell$=3 component, combined with the deduced ground state branching, allows to indirectly determine the partial cross section for this elusive 7/2$^-$ state.

A possible indication of the location of the 7/2$^-$ state is provided by the comparison of the populated states in \ce{^33Mg} from the one-neutron and one-proton removal reactions. The differences observed in the one-proton removal reaction can reveal the composition of the states not populated in that reaction. The most important difference is the disappearance of both 549 keV and 703 keV transitions. Since the removal of a $d_{5/2}$ proton from \ce{^34Al} is expected to populate mainly positive parity states in \ce{^34Al}, the disappearance of both transitions could indicate that the 549 keV transitions corresponds to the population of the unobserved 7/2$^-$ isomeric state from the 703 keV state, which would place this state at around 154 keV, reaching a similar conclusion to the work in \cite{NUM01} where the presence of this isomer was also inferred.

Partial cross sections are compared to theoretical ones based on the eikonal reaction model and shell model calculations using 3 modern interactions (IMSRG, SDPF-MU and SDPF-U-SI). They reveal the complicated nature of the states in \ce{^33Mg} where large configuration mixing with p-h excitations play a very important role. Both the IMSRG and SDPF-MU reproduce the observed cross sections rather well, especially when taking into account the strength feeding the indirectly observed 7/2$^-$ state at 154 keV.

Clearly additional data is needed on the spectroscopy of this nucleus. The one-neutron removal reaction from $^{34}$Mg would be advantageously performed using a high-resolution $\gamma$-ray tracking array, in order to better resolve the numerous transitions and $\gamma$-$\gamma$ coincidences. However, the non-observation of the prompt decay from the inferred isomeric 7/2$^-$ state would remain in such an improved experiment. From the estimates of shell model calculations, the delayed 154 keV $\gamma$-ray from the decay of the isomer could be detected in a setup with $\gamma$-ray detection around the implantation site of the \ce{^33Mg} residues. A direct search for this isomer could also be performed from a decay experiment using a beam of \ce{^33Mg} produced via projectile fragmentation. Another possibly more promising venue would be using a (d,p) transfer reaction on a $^{32}$Mg beam in inverse kinematics, that would be able to measure and identify the $\ell$=3 strength directly, as well as to other populated states. The parity of the 484 keV state in particular, could not be determined in this experiment, but would be easily identified in a transfer reaction study. Such a \ce{^32Mg} beam will be available in the early days of FRIB operations, produced via projectile fragmentation and re-accelerated by the ReA6 linac to energies close to 10 MeV/u.

\section{Acknowledgements}
The authors would like to thank the RIKEN Nishina RIBF accelerator staff and the BigRIPS team for providing the high intensity \ce{^48Ca} primary beam and the resulting high quality \ce{^34Mg} and \ce{^34Al} radioactive beams. DB acknowledges fruitful discussions with A. Gade. This work is supported in part by the U.S. National Science Foundation under grant PHY-0606007 and the U.S. Department of Energy, Office of Science, Office of Nuclear Physics under grant DE-SC0020451. SRS is supported by the U.~S. Department of Energy under Contract DE-FG02-97ER41014.

\bibliography{biblio.bib}

\end{document}